\documentstyle[12pt]{article}

\def\endpage{\vfill\eject}

\newcommand{\AmS}{{\protect\the\textfont2
\renewcommand{\thesection}{\Roman{section}}
  A\kern-.1667em\lower.5ex\hbox{M}\kern-.125emS}}
\begin{document}
\rightline {\bf DFTUZ/96/3}
\vskip 2. truecm
\centerline{\bf TESTING LOGARITHMIC VIOLATIONS TO SCALING}
\centerline{\bf IN STRONGLY COUPLED QED}
\vskip 2 truecm
\centerline { V.~Azcoiti$^a$, G. Di Carlo$^b$, A. Galante$^{c,b}$, 
A.F. Grillo$^d$, V. Laliena$^a$ and C.E. Piedrafita$^b$}
\vskip 1 truecm
\centerline {\it $^a$ Departamento de F\'\i sica Te\'orica, Facultad 
de Ciencias, Universidad de Zaragoza,}
\centerline {\it 50009 Zaragoza (Spain).}
\vskip 0.15 truecm
\centerline {\it $^b$ Istituto Nazionale di Fisica Nucleare, 
Laboratori Nazionali di Frascati,}
\centerline {\it P.O.B. 13 - Frascati 00044 (Italy). }
\vskip 0.15 truecm
\centerline {\it $^c$ Dipartimento di Fisica dell'Universit\'a 
dell'Aquila, L'Aquila 67100 (Italy)}
\vskip 0.15 truecm
\centerline {\it $^d$ Istituto Nazionale di Fisica Nucleare, 
Laboratori Nazionali del Gran Sasso,}
\centerline {\it Assergi (L'Aquila) 67010 (Italy). }
\vskip 3 truecm
\centerline {ABSTRACT}
Using very precise measurements of the critical couplings for the chiral 
transition of non compact $QED_4$ with up to 8 flavours, 
we analyse the behaviour of the order 
parameter at the critical point using the equation of state of a 
logarithmically improved scalar mean field theory, that of the
Nambu-Jona Lasinio theory and a pure power law. 
The first case is definitively excluded by the 
numerical data. The stability of the fits for the last two cases, as well 
as the behaviour with the number of flavours 
of the exponent of the logarithmic violations to the 
scaling favour clearly a pure power law 
scaling with non mean field exponents.

\vfill\eject

\par
Non perturbative renormalizable gauge theories are among the most stimulating
and less known subjects in theoretical physics. It is not known if it 
is possible 
to construct such a theory in four dimensions and the standard prejudice is 
that only asymptotically free gauge theories with gaussian fixed points are 
"good" field theories.

Non compact $QED_4$ or its generalized version, the gauged Nambu-Jona Lasinio 
$(GNJL)$ model, are good candidates to analyze this problem. Both models have 
the common feature that they couple fermions strongly enough to produce 
fermion condensates, and therefore a phase, where chiral symmetry is 
spontaneously broken, appears at sufficiently strong gauge coupling. Since 
composite scalars are present in the spectrum of these models, the existence 
of a non trivial continuum limit is strongly related, as 
discussed in \cite{SACHA}, with the balance between the fermion 
attraction due to the interaction and the zero point repulsion due to the 
kinetic energy. If the short distance attraction is too strong, composite 
scalars with vanishing physical size appear in the spectrum, thus giving rise 
to a non interacting field theory. Even if no rigorous proof exists, this 
seems to be the case in the Nambu-Jona Lasinio $(NJL)$ model where mean field 
exponents with logarithmic violations to scaling driving to a vanishing 
renormalized coupling are expected \cite{KKK}.

In a recent investigation of the gauged $NJL$ model \cite{NOSOTROS} 
we have found strong evidence 
supporting the fact that the gauge interaction can change qualitatively 
the trivial scenario of the $NJL$ model. Stimulated by these results and 
following our investigation on non compact $QED_4$ 
started several years ago \cite{IJMP}, 
we want to report in this letter some new results concerning the critical 
behaviour of this last model. 

Our analysis is based on very precise determinations 
of the critical couplings obtained from the computation of the 
susceptibilities 
in the Coulomb phase \cite{SUSCEP} on $10^4, 12^4$ and $14^4$ lattices. From 
these results we analyse the behaviour of the order parameter at the critical 
point and for several number of flavours
by fitting it with three different equations of state $(EOS)$: i) The 
$EOS$ of a logarithmically improved scalar mean field theory, ii) a power 
law scaling without logarithmic violations and iii) a $NJL$ 
model-like $EOS$.

The case i), as will be shown in what follows, is definitively excluded by 
the numerical data. This is not surprising at all since as suggested in 
\cite{SACHA} and corroborated in \cite{KKK}, triviality 
in a theory with composite scalars should manifest 
in a different way than in a theory with fundamental scalars. In fact the 
logarithmic violations to scaling in the first case, as in the $NJL$ model, 
are expected to have an effective $\delta$ exponent less than 3 whereas in 
the second case an effective $\delta$ larger than 3 is obtained.

The difficult task is to distinguish from the numerical data between ii) 
(power law scaling) and iii) (four fermion $EOS$) since in both cases the 
effective $\delta$ is less than three.  
Let us anticipate that even if the behavior of the order parameter 
at the critical point is well fitted by both equations of state, 
the stability of the fits with lattice size and the behavior of the exponent 
which controls the logarithmic violations to scaling with the 
number of flavours strongly favour a pure power law scaling 
against logarithmic violations to mean field.

The main ingredient of this analysis is the very precise determination 
of the critical coupling $\beta_c$ in $QED_4$ which follows from the 
computation of the susceptibility in the Coulomb phase of this model. 
A detailed explanation of this computation for lattices up to $10^4$ 
can be found in \cite{SUSCEP}. 

Let us recall the two essential ingredients we have used in this analysis.

First, exploiting the potentialities of the Microcanonical Fermionic Average 
($MFA$) approach to simulate 
dynamical fermions in gauge theories \cite{MFA} we have done simulations 
in the chiral limit at exactly zero fermion mass, which allows to overcome 
the ambiguities in the fermion mass extrapolations. 

Second, we have also 
exploited the fact that in the Coulomb phase of this model, characterized 
by a non degenerate ground state, we can interchange the chiral limit 
with the thermodynamical limit. Notice that the reason why this 
exchange is not allowed in general cases ($QCD$ simulations, broken 
phase of $QED$, etc.) is the degeneration of the vacuum state. If we are 
in a broken phase where many degenerate vacua connected by symmetry 
transformations coexist, a permutation of the chiral and thermodynamic 
limits would imply a path integral over all the Gibbs state. Then several 
unpleasant facts like the violation of the cluster property for correlation 
functions, vanishing values for the order parameter or wrong values for 
the computed susceptibilities will appear. However if calculations are 
done in the symmetric phase characterized by a non degenerate ground state, 
there are not, in general cases, physical reasons to prevent from doing 
such a permutation.

The most general expression for the vacuum expectation value of any 
operator $O$, after integration of the Grassmann variables, can be 
written as \cite{MFA}

$$<O> = {\int dE n(E){\overline{O\det}\Delta\over{\overline{\det}\Delta}}
 e^{-6V\beta E-S^{F}_{eff}(E,m)} 
\over {\int dE n(E) e^{-6V\beta E-S^{F}_{eff}(E,m)}}} 
\eqno(1)$$

\noindent
where 

$$n(E)=\int [d A_\mu] \delta(6VE-S_G[A_\mu])
\eqno(2)$$

\noindent
is the density of states at fixed energy $E$ and
$S^{F}_{eff}(E,m)$ in (1) is the fermion effective action defined 
as 

$$S^{F}_{eff}(E,m) = -\log \overline{\det} \Delta(E,m)
\eqno(3)$$

\noindent
$\overline{O\det}\Delta$ means the mean value of the product of 
the operator $O$ times the fermionic determinant, computed over gauge 
field configurations at fixed energy $E$.

Since we are interested here in the computation of susceptibilities in 
the chiral limit, we 
will use the previous expression for the particular cases in which $O$ 
is the longitudinal or transverse susceptibility.

The longitudinal and 
transverse susceptibilities in the Coulomb phase are equal except a sign.
They can be 
computed by taking for the operator $O$ the expression

$$O={2\over V} \sum_i  {1\over{\lambda^2_i}}
\eqno(4)$$

\noindent
where the sum in (4) runs over all positive eigenvalues of the massless 
Dirac operator.

The transverse susceptibility $\chi_T$ in the broken phase diverges always 
because of the Goldstone boson. The longitudinal susceptibility $\chi_L$ 
on the other hand, can not be computed in the broken phase using equation 
(4) since in this phase the ground state is not invariant under chiral 
transformations. 

We refer the reader interested to details on the computation of 
susceptibilities to \cite{SUSCEP}. Let us recall that using the $MFA$ 
approach, computations at several number of flavours $N_f$ can be done 
without extra computer cost \cite{MFA}. 
In table I we report the critical couplings 
extracted from the susceptibilities in $10^4,12^4$ and $14^4$ lattices 
at $N_f=1,2,3,4,8$. The most striking fact of this table is the small 
statistical errors of the critical couplings which follows from the 
high quality of the susceptibility fits, as reported in \cite{SUSCEP}.

We have fitted  the behaviour of the chiral condensate 
$\langle\,\bar\chi\chi\,\rangle$ against the bare fermion mass $m$ at the 
critical values of $\beta$ reported in Table I using different equations 
of state. In Fig. 1 we plot 
${\langle\,\bar\chi\chi\,\rangle^3}
/{log\langle\,\bar\chi\chi\,\rangle}$ 
against $m$ in the mass interval $(0.005,0.1)$ 
for the four flavour theory and 
in a $14^4$ lattice. Were the critical behaviour of this model described 
by a gaussian fixed point as in a logarithmically improved scalar mean field 
theory, the points in the small mass region of this figure should be 
well fitted by a straight line crossing the origin. The solid line in this 
figure is the best fit obtained under the previous assumption. The very bad 
quality of this fit $({\chi^{2}\over{d.o.f.}}=1267, m\le 0.1, 
{\chi^{2}\over{d.o.f.}}=358, m\le 0.05)$ 
disproves definitively 
such a possibility. Similar results have been obtained in smaller lattices 
and for different flavour numbers.

Having shown that the critical behaviour of non compact $QED$ can not 
be described by a logarithmically improved scalar mean field theory, we 
will try the same kind of fit for the other two reasonable possibilities, 
i.e. pure power law behaviour and mean field with logarithmic violations 
a la Nambu-Jona Lasinio.

Fig. 2 is a plot of the same data reported in Fig. 1 but in the ordinate 
axis we plot $\langle\,\bar\chi\chi\,\rangle^{2.80}$. The solid line is 
the best fit of all the points at small $m$ with a straight line crossing the 
origin. The high quality of this fit is corroborated by the value 
${\chi^{2}\over{d.o.f.}}=0.38$, a value which is stable until masses of the 
order of $0.08$ and increases slowly when masses larger than $0.08$ are 
included in the fit. 
The value 2.80 of the $\delta$ exponent 
has been chosen as the best one for the linear fit (see Table II and the 
discussion on the determination of $\delta$ at the end). From the results 
reported in Fig. 2 we conclude that a pure power with 
$\delta=2.80$ describes with high accuracy the behaviour of the chiral 
condensate at the critical coupling.

Our last plot for the chiral condensate at the critical coupling is 
reported in Fig.3. For this plot we use a equation of state a la Nambu-Jona 
Lasinio

$${\langle\,\bar\chi\chi\,\rangle^3}{log^p({1\over{
\langle\,\bar\chi\chi\,\rangle}})} = C m
\eqno(5)$$

\noindent
where $C$ in (5) is a constant and the exponent $p$ of the logarithmic 
violations to scaling is 
left as a parameter of the fit. Recall that $p=1$ in the large $N_f$ 
limit and that this result does not changes after taking into account 
the $1\over N_f$ \cite{SACHAKOGUT} and $1\over {N_f^2}$ \cite{GRACEY} 
corrections, this suggesting that a value of $p$ different from $1$ is 
not very reliable. Notwithstanding that we decided to left 
$p$ as a parameter of the fit for two reasons: i) our results for the 
chiral condensate does not support a fit with equation (1) and $p=1$ 
and ii) there is no rigorous proof that $p$ does not changes with $N_f$. 

The best fit of our results in the four flavour model 
with equation (5) is obtained for $p=0.28$. In Fig. 3 we plot 
${\langle\,\bar\chi\chi\,\rangle^3}{log^{0.28}({1\over{
\langle\,\bar\chi\chi\,\rangle}})}$
against $m$. The solid line in this 
figure is a fit of all the points at small $m$ with a straight line 
crossing the origin. Again now, as in the case of the power law behaviour, 
we get a high quality fit of these points $({\chi^{2}\over{d.o.f.}}
=0.89)$. 

At first sight it seems difficult to decide between the last two cases 
(power law and Nambu-Jona Lasinio). However we can get some insight on the 
reliability of the last two fits by analyzing 
their stability with the lattice 
size as well as the flavour dependence of the $p$ 
exponent in equation (5).

Table II contains a summary of the results we obtain for the $\delta$ 
and $p$ exponents for 
the three lattice sizes $10, 12, 14$ and $N_f=1,2,3,4,6,8$. The values 
of $\delta$ reported in this table are obtained fitting the chiral 
condensate 
with a pure power law (case ii previously, corresponding to non mean-field 
exponents). Using instead 
a Nambu-Jona Lasinio like $EOS$, where $\delta=3$, we obtain 
values of the $p$ exponent of equation (5), which we also
report in Table II. The errors in this Table 
take into account both, the errors of the fits and the error in the 
determination of the critical couplings (see Table I). The high precision 
of the exponents reported in Table II follows from the very small errors 
in the critical couplings of Table I.

Looking at the results reported in Table II we can discuss both, the 
stability of the results with the lattice size and the flavour dependence 
of $p$. First notice that the values of $\delta$ for the three 
lattice sizes at each value of $N_f$ are always compatible whereas 
this does not hold for the values of $p$. We conclude that 
the pure power law fits are much more stable with lattice size than 
$NJL$-like fits. 

The second important fact that can be observed in this 
Table is that the value of the $p$ exponent, contrary to expectations 
based on the $1\over N_f$ expansion of the $NJL$ model, not only is 
different from $1$ but 
decreases when $N_f$ increases and 
therefore goes away from the expected $p=1$ at large $N_f$. As well 
known \cite{SACHAETAL},\cite{LARGEN} the chiral 
transition of noncompact $QED$ changes from second to first order at 
large $N_f$ and therefore we can not extrapolate our results for $p$ 
in Table II to $N_f=\infty$. Notwithstanding that, Kim, Kocic and Kogut 
have shown in \cite{KKK} how numerical simulations of the $NJL$ 
model with discrete $Z_2$ symmetry at $N_f=12$ reproduce very well 
the logarithmic violations to mean field scaling given by the $EOS$ 
of the model at $N_f=\infty$. Therefore, were the critical behaviour 
of non compact $QED$ described by the $NJL$ model, we would expect a 
value of $p\sim 1$, at least at $N_f=8$.

In conclusion, we believe that the features previously discussed 
strongly favour a pure power law scaling with 
non mean field exponents against logarithmic violations to mean field 
a la Nambu-Jona Lasinio. This is a very interesting and important 
improvement with 
respect to previous work on this subject \cite{IJMP},\cite{CUATRO} and 
it has been possible as a consequence of the very precise measurements of the 
critical couplings reported in Table I.

We wish to acknowledge the discussions with A. Koci\'c. 
This work has been partly supported through a CICYT (Spain) - 
INFN (Italy)
collaboration.

\endpage

%#
\endpage
\vskip 1 truecm
\leftline{\bf Figure captions}
\vskip 1 truecm
{\bf Figure 1.} Numerical results for 
${\langle\,\bar\chi\chi\,\rangle^3}
/{log\langle\,\bar\chi\chi\,\rangle}$ 
against $m$ 
for the four flavour theory and 
in a $14^4$ lattice. The solid line in this 
figure is the best linear fit crossing the origin. 

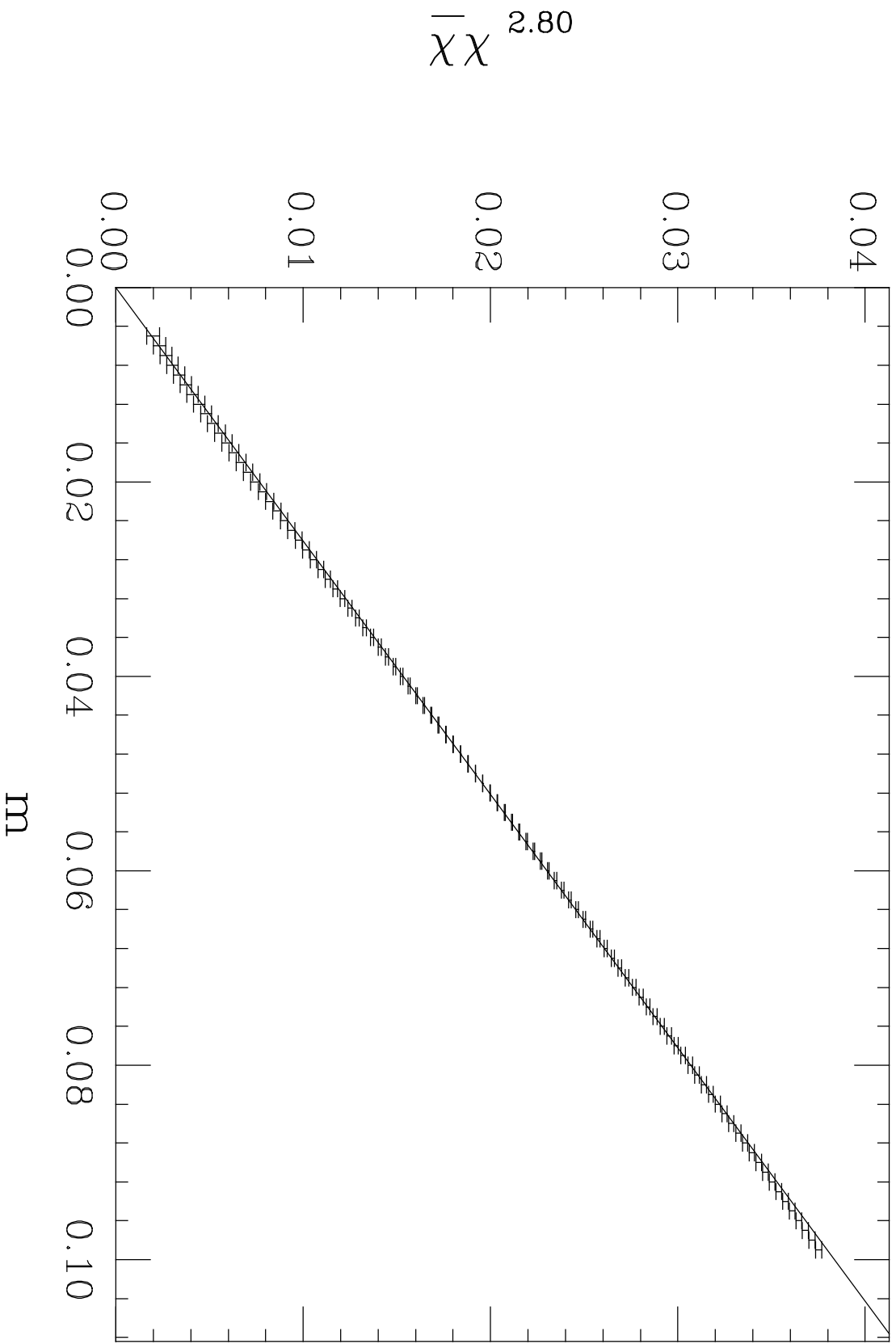
{\bf Figure 2.} Plot of 
$\langle\,\bar\chi\chi\,\rangle^{2.80}$ against $m$. The solid line is 
the best fit of all the points at small $m$ with a straight line crossing the 
origin. 

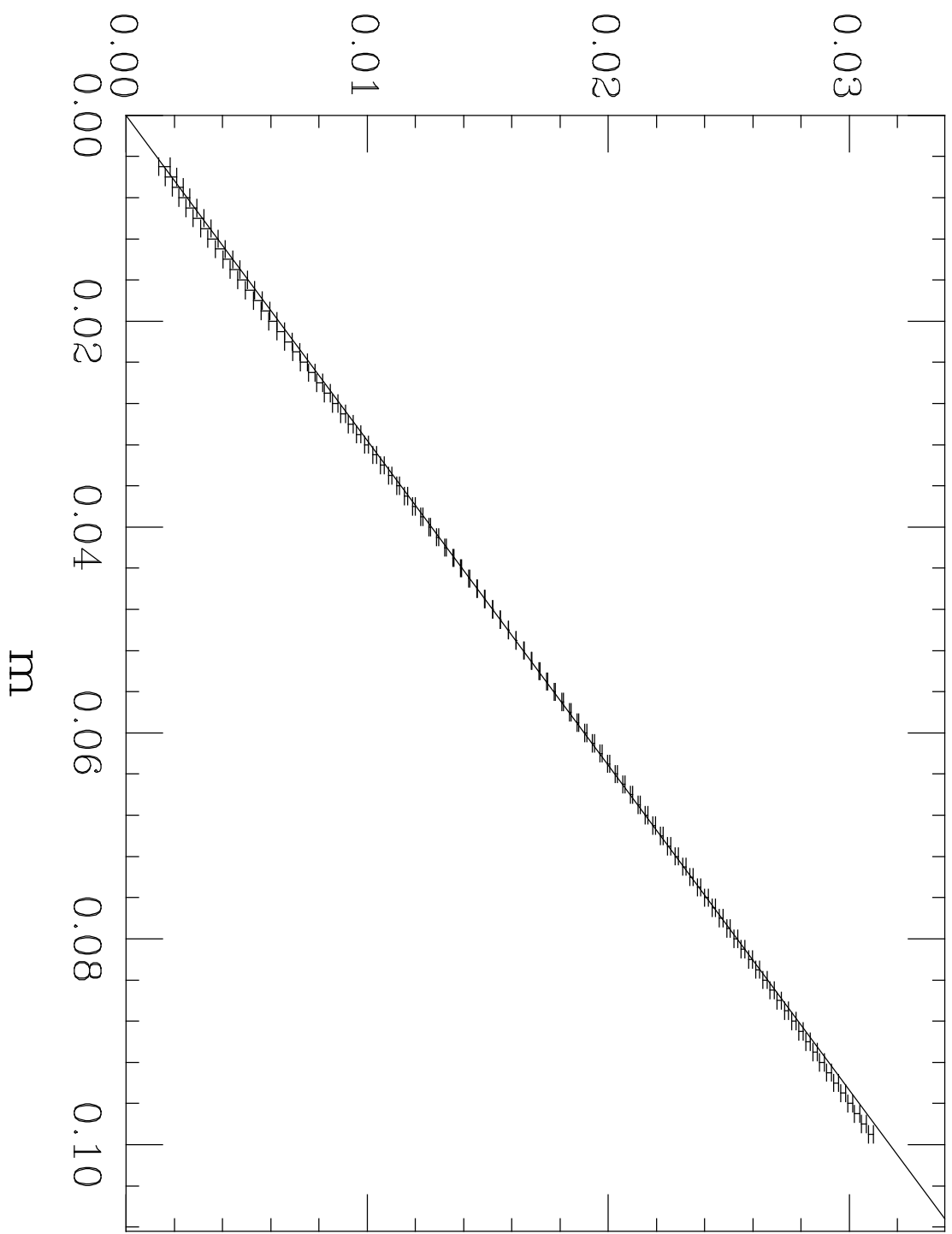
{\bf Figure 3.} $ \langle\,\bar\chi\chi\,\rangle^3 log^{0.28}({1\over
\langle\,\bar\chi\chi\,\rangle}) $ against $m$. The solid line 
is a fit of all the points at small $m$ with a straight line 
crossing the origin.

\endpage
\vskip 1 truecm
\leftline{\bf Table captions}
\vskip 1 truecm
{\bf Table I} Critical couplings 
extracted from the susceptibilities in $10^4,12^4$ and $14^4$ lattices 
at $N_f=1,2,3,4,8$.

{\bf Table II} Results for the $\delta$ and $p$ exponents for 
the three lattice sizes $10, 12, 14$ and $N_f=1,2,3,4,6,8$. The values 
of $\delta$ reported in this table correspond to the results for the fits 
of the chiral condensate 
with a pure power law whereas in the case of the 
Nambu-Jona Lasinio like $EOS$, where $\delta=3$, we report 
the value of the $p$ exponent in equation (5).

\end{document}